\documentclass[12pt, nofootinbib]{article}

\usepackage{graphicx}
\def\eg{{\it e.g.}}

\renewcommand{\title}[1]{\vbox{\center\LARGE{#1}}\vspace{5mm}}
\renewcommand{\author}[1]{\vbox{\center#1}\vspace{5mm}}
\newcommand{\address}[1]{\vbox{\center\em#1}}

\renewcommand{\date}[1]{\vbox{\center#1}}

\def\gU{\textsf{U}}

\def\grad{\vec \nabla}

\usepackage[usenames,dvipsnames]{xcolor}
\usepackage[pdftex, bookmarks={false}, pdfauthor={John McGreevy}, pdftitle={This is a secret message!}]{hyperref}
\hypersetup{colorlinks=true, linkcolor=BrickRed, citecolor=Violet, filecolor=OliveGreen, urlcolor=RoyalBlue, filebordercolor={.8 .8 1}, urlbordercolor={.8 .8 0}}

\begin{document}

\title{Viewpoint: duality for fermionic vortices}

\author{John McGreevy}

\address{Department of Physics\\
University of California at San Diego\\
{\tt mcgreevy@physics.ucsd.edu}}

\date{\today}

\begin{abstract}
This is a 
version (less professional editing, more details and jokes) of an 
\htmladdnormallink{APS Physics Viewpoint}{http://physics.aps.org/} 
 about three recent papers extending charge-vortex duality to 
 fermionic vortices \cite{Metlitski:2015eka, Wang:2015qmt, 2015arXiv151008455M}.
\end{abstract}

{\bf Duality.}  Humans are not very good
at understanding interacting many body systems.
We understand 
the behavior of free particles, 
and can sometimes use perturbation theory to incorporate weak interactions between them.
But what happens when we couple the particles strongly?
We have had to rely largely on experiments to suggest the many interesting collective phenomena 
that can result\footnote{Examples include: superconductivity, superfluidity, 
the quantum Hall effects, fractionalization of charge and spin, the wetness of water...}.

{\it Duality} is the circumstance
in which cranking up the 
interactions results 
in another system we understand, 
usually described in terms of 
another set of weakly-coupled particles.
You might think this is a failure of imagination (we thought we had 
two models but really they are the same), 
or you might think this is evidence that we are doing a good 
job thinking of all the possibilities (this is probably wrong 
given our track record).
Either way, when this happens, it is a solution to the strong coupling problem.  A further reason that theoretical physicists get excited about duality is that it undermines the very notion of an `elementary particle'.  In terms of the original variables, the dual weakly coupled particles are generally {\it solitons} -- large, heavy classical objects made up of many quanta.  The amazing thing is that solitons can become `elementary,' in the sense that the model can be reformulated with them as the basic constituents.  An example of a duality, which we will use below, 
arises in Maxwell's theory of electromagnetism,
where the duality exchanges 
electric and magnetic point charges, 
and relates a theory with interaction strength $e$ to
one with interaction strength $1/e$.
This idea has been generalized to many other field theories,
particularly with tools of supersymmetry\footnote{For more examples and references, I recommend \cite{Polchinski:2014mva}.
Evidence can be found for the duality symmetry 
of (3+1)-dimensional abelian gauge theory
in \cite{Cardy:1981qy, Cardy:1981fd} (on the lattice)
and in \cite{Witten:1995gf, Verlinde:1995mz} (in the continuum, at energies below the electron mass).}.

{\bf Classic duality for vortices.}
A {\it vortex} 
is a localized object in the plane around which a phase field 
(a field that takes values on the circle)
winds. 
Dualities relating particles and vortices have a rich history
in particle physics and condensed matter physics.
The simplest version relates 
the particles in the normal state of a superfluid to the vortices in a dual superconductor
\cite{Banks:1977cc, Peskin:1977kp, Thomas:1978aa, PhysRevLett.47.1556, PhysRevB.39.2756}.
Specifically, it maps the $2+1$-dimensional XY model
-- a model with a $\gU(1)$ symmetry, such as particle number --
 (near its Wilson-Fisher critical point),
to the abelian Higgs model 
-- the low-energy description of a superconductor --
(near its critical point).
The density of particles maps to the density of magnetic flux\footnote{The gauge field $\vec a$ can be regarded as a 
solution of the continuity equation $ \grad \cdot \vec J=0$ for 
the particle 3-current:
$ \vec J \propto  \grad \times \vec a $.}.
In the symmetry-broken phase, the Goldstone boson is the photon
(which has only one polarization state in 2+1d).
The symmetric (Mott insulator) phase is achieved
when the proliferation of vortices disorders the condensate of bosons;
the resulting gapped state is the Higgs phase of the gauge theory, 
where the photon eats the phase field and becomes massive.
This charge-vortex duality
has 
been a workhorse in the study of two dimensional condensed matter systems,
such as 
fractional quantum Hall states \cite{PhysRevLett.63.903}, 
superconducting thin films \cite{PhysRevB.39.2756, FG9087}, 
and
beyond-Landau critical phenomena \cite{2004PhRvB..70n4407S}.

{\bf Fermionic vortices.} 
Vortices can be fermions\footnote{It is not so easy to give a simple example where this happens.  
In type II superconductors,
a vortex supports bound fermion modes \cite{CAROLI1964307}.
The resulting boundstate of a quasiparticle and a vortex
is, however, still a boson; 
this is because the charge and magnetic flux 
have a long-ranged statistical interaction symptomatic of topological order.
So these two exotic phenomena (fermionic modes bound to the vortices and topological order) cancel each other out, 
and the ordinary bosonic particle-vortex duality applies
even in this case.}.
A dual description of such a system
requires a new ingredient, since it  
must include a fermionic particle.
An appealing proposal for the required duality 
has been made by Metlitski and Vishwanath \cite{Metlitski:2015eka} 
and independently by Wang and Senthil \cite{Wang:2015qmt}
(building on earlier work of Son \cite{Son:2015xqa}),
and further illuminated
by Mross, Alicea and Motrunich \cite{2015arXiv151008455M}:
the fermionic vortex of QED in 2+1 dimensions is a free Dirac fermion.
This deceptively simple statement 
joins together 
several 
disparate lines of progress in theoretical physics: 
interacting topological insulators, anomalies in quantum field theory,
spin liquids, compressible quantum Hall states.

{\bf Interacting topological insulators.} 
Topological insulators (TIs) (and more generally symmetry-protected topological states)
are nearly-trivial states of matter, distinguished only by their edge states\footnote{For a review of interacting TIs, see \cite{Senthil:2014ooa}.}.
The most famous example is a (3+1)-dimensional time-reversal invariant 
TI, which can host an odd number of Dirac cones (such as one) at 
the boundary of the sample.
A time-reversal-symmetric realization of a single two-dimensional Dirac cone
is not possible without the bulk TI, 
so it provides a signature of the bulk phase.
 But the Dirac cone is not the only possible signature.
Deformations of the boundary conditions 
can dramatically change the spectrum of the edge states, 
but they preserve some essential ``TI-ness";
this essence is called an {\it anomaly} in the high energy theory literature.


{\bf Making use of the bulk.}  The argument of \cite{Metlitski:2015eka, Wang:2015qmt}
uses this ambiguity in the edge theory, 
and the electric-magnetic duality of an auxiliary 
theory of bulk electromagnetism coupled to the TI.
When I said above that (3+1)-dimensional QED had a duality that 
interchanged the electric and magnetic charges, 
you could have complained that the electron is a fermion, while a magnetic monopole may not be.
However, the authors observe that in a 
TI, a charge-two monopole is also a fermion
\cite{Metlitski:2015eka, Wang:2015qmt}.
By analyzing the spectrum of electric and magnetic charges, 
they conclude that the duality exchanges the electron and this double monopole, 
and moreover exchanges a TI with time-reversal symmetry
with a `chiral TI'\footnote{That is, it exchanges Class AII and AIII of the tenfold-way classification of TIs \cite{Senthil:2014ooa}.}
which instead has particle-hole symmetry.  
But this is just what is required of a symmetry  which
exchanges a 
fermionic particle with a fermionic vortex: 
time reversal symmetry maps the electron to itself,
but the vortex to a vortex winding the other way.
The exchange of particles and holes, in turn, preserves the vortex.
The bulk duality\footnote{Further strong mathematical evidence for the duality relating 
electromagnetism coupled to 
class AII and class AIII
TIs was provided in a detailed calculation by 
\cite{Metlitski:2015yqa}.
} implies a duality between edge theories.

\begin{center}
\includegraphics[width=\textwidth]{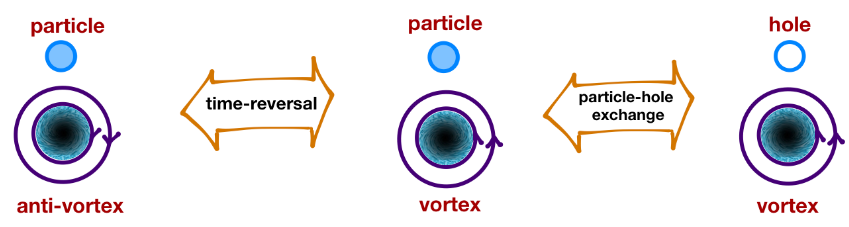}
{\footnotesize Figure: The action of discrete symmetries on the fermionic particle (top) and 
its dual fermionic vortex (bottom).}
\end{center}


The freedom to deform the edge of the TI 
implies a weak notion of equivalence: 
although the edge theories are guaranteed to be the same 
in terms of topology (charge sectors, symmetry assignments, anomalies), 
they need not have identical low-energy physics.
One side of the duality is a free, massless Dirac particle, 
but the long-distance physics of (2+1)-dimensional
QED is a long-standing problem, 
which is important in the study of spin liquids\footnote{See \eg~\cite{PhysRevB.70.214437} 
and for recent 
progress, see \cite{Grover:2012sp}.}.
Abelian gauge theory in two dimensions tends to 
develop an energy gap
because of the proliferation of monopoles \cite{Polyakov:1976fu}.
In the realization proposed in \cite{Metlitski:2015eka, Wang:2015qmt}, 
this cannot happen because those monopoles carry electric charge, 
and indeed the stronger notion of duality as
an equivalence of low-energy physics may hold.

{\bf Dimensional bootstrap.}  At a more microscopic level,
a lattice version of the duality 
was derived in \cite{2015arXiv151008455M}.
Two aspects of this construction are notable:
First, it relies on an 
explicit change of variables, 
which (like the Jordan-Wigner transformation which makes fermions 
out of spins in one dimension\footnote{For a pedagogical treatment I will recommend 
\S2.2.5 of 
\htmladdnormallink{these lecture notes}{http://physics.ucsd.edu/~mcgreevy/s14/239a-lectures.pdf}.}) can be explicitly checked,
but remains arcane.
Second, the lattice models in the construction are 
made out of systems that do not themselves exist
except at the edge of a TI, namely an array of wires carrying chiral fermions.
The derivation on the lattice provides evidence for the strong form of the duality, but 
is not a proof since
a phase transition may intervene between the lattice model
and the strongly-coupled continuum limit.

{\bf Impact.} Furious activity surrounds this development.
It immediately solves outstanding puzzles about interacting TIs 
and their symmetric, gapped topologically-ordered surface states
\cite{Metlitski:2015eka, Wang:2015qmt, Wang:2016fql}.
Moreover, 
the `auxiliary' bulk gauge theory 
in the above discussion 
actually describes a novel 
3d spin liquid with time-reversal symmetry \cite{2016PhRvX...6a1034W, Wang:2016fql}, 
which was in fact the starting point for the authors of \cite{Wang:2015qmt}.
Pyrochlore magnets made from rare-earth elements\footnote{four of seventeen of which are named after the same village of Ytterby, Sweden
where they were discovered}
provide candidate materials for such phases.

Perhaps most interestingly, 
ingredients of this work grew out of the attempt
to construct a manifestly particle-hole symmetric theory of
a half-filled Landau level
\cite{Son:2015xqa}, 
the best-understood non-Fermi liquid state.
Briefly, the successful theory of this state \cite{HL9312} 
makes a distinction between half-filled and half-empty\footnote{This brilliant joke 
provides the title of \cite{Murthy:2016jnc}.}.
The associated particle-hole symmetry is emergent, 
and an exact local realization can only occur at the edge of a TI.
The main outcome is a new understanding of
the Dirac-like nature of the composite fermion excitation of this system, 
which provides a new point of view on the phenomenology 
\cite{Son:2015xqa, Wang:2016fql, 2015arXiv150804140G, 2015arXiv151206852P,2016arXiv160406807W}\footnote{See 
\cite{alicea-commentary:2015}
for commentary on the connection with the half-filled Landau level.}.
This circle of ideas has
also 
led to new insights on 2d superfluid-insulator 
transitions with nearby metallic phases, where the fermions come alive \cite{Mulligan:2015zua, Mulligan:2016rno}.
More recently, consequences of the fermionic vortex duality have been checked mathematically \cite{Metlitski:2015yqa}
and numerically \cite{2015arXiv150804140G},
and the duality has been fit into the wider web of field theory dualities
\cite{Karch-Tong-2016, Murugan-Nastase-2016, TheEndOfTheWorld}\footnote{In particular, \cite{TheEndOfTheWorld}
gives a clear account of the subtle global issues 
in the duality, 
and explains a previously-mysterious action of time-reversal symmetry
on a description of a superfluid
in terms of fermions in Chern bands
\cite{PhysRevB.89.235116}.}.

{\bf Acknowledgements.}
I am grateful to S.~M.~Kravec and T.~Senthil for helpful comments,
to Matteo Rini and Jessica Thomas for editorial suggestions,
and to Ashvin Vishwanath for correcting a misconception in an earlier version.
My work is supported in part by
funds provided by the U.S. Department of Energy
(D.O.E.) under cooperative research agreement 
DE-SC0009919.


\bibliographystyle{ucsd}
\bibliography{collection} 
\end{document}